\documentclass[english]{DISproc}
\usepackage[T1]{fontenc}
\usepackage[latin9]{inputenc}
\usepackage{wrapfig}
\usepackage{graphicx}

\makeatletter

\providecommand{\tabularnewline}{\\}

\newlength{\mytextwidth}
\makeatother

\usepackage{babel}

\begin{document}

\title{Progress in CTEQ-TEA PDF analysis}

\author{\textsl{Pavel Nadolsky},\textsl{$^{a}$}%
\thanks{Presenter%
} \textsl{Jun Gao,$^{a}$} \textsl{Marco Guzzi,$^{a}$} \textsl{Joey
Huston,$^{b}$} \textsl{Hung-Liang Lai,$^{c}$} \textsl{Zhao Li,$^{b}$}
\textsl{}\\
\textsl{Jon Pumplin,$^{b}$} \textsl{Dan Stump,$^{b}$} \textsl{C.-P.
Yuan$^{b,d}$} \\
 $^{a}$Department of Physics, Southern Methodist University, Dallas,
TX 75275, USA\\
 $^{b}$Department of Physics and Astronomy, Michigan State University,
E. Lansing, MI 48824, USA\\
 $^{c}$Taipei Municipal University of Education, Taipei, Taiwan
\\
 $^{d}$Center for High Energy Physics, Peking University, Beijing,
China }
\maketitle
\begin{abstract}
Recent developments in the CTEQ-TEA global QCD analysis are presented.
The parton distribution functions CT10-NNLO are described, constructed
by comparing data from many experiments to NNLO approximations of
QCD. 
\end{abstract}
The global analysis of QCD makes use of experimental data from many
short-distance scattering processes to construct, within some approximations,
universal parton distribution functions (PDFs) for the proton. Then
these functions can be used to calculate hadronic cross sections in
the Standard Model and other theories. Global analysis and the resulting
PDFs are necessary for the interpretation of experimental results
at hadron colliders.

Recently published PDFs are based on next-to-next-to-leading order
(NNLO) approximations for perturbative QCD\,\cite{NNLOPDFs}. Complete
calculations for this order of approximation are available for the
running coupling $\alpha_{{\rm s}}(Q)$, PDF evolution in $Q$, matrix
elements in deep-inelastic scattering \cite{NNLODIS} and vector boson
production \cite{NNLOVBP}. The CTEQ analysis treats quark-mass effects
in the S-ACOT-$\chi$ factorization scheme, which has been recently
extended to two-loop, or NNLO, accuracy~\cite{CTEQ:11}. Though the
NNLO matrix elements are still unknown for some important processes,
such as the inclusive jet production in $pp/p\overline{p}$ collisions,
it is important to use NNLO approximations, where available.

CTEQ has developed PDFs for general-purpose computations and estimates
of PDF-driven uncertainties over many years \cite{PrevCTEQ}. The
most recent PDFs in this class, named CT10 and CT10W, were published
in 2010 \cite{CT10}. We now present a new family of CTEQ parton distributions,
named CT10 NNLO. There are several reasons for publishing them. First,
the CT10 NNLO global analysis is based on the NNLO approximation of
perturbative QCD, whereas the CT10 and earlier analyses were based
on NLO. Second, benchmarking of NLO jet cross sections \cite{Maestre:2012vp}and
DIS cross sections was performed to quantify theoretical uncertainties,
and an in-depth study of the treatment of correlated experimental
errors has been completed. Third, selection of experimental data sets
has been revisited. The new NNLO PDFs are closely related to both
CT10 and CT10W NLO PDFs and can be matched to either of two NLO PDF
sets when comparing the NLO and NNLO cross sections. In all three
cases, only data from pre-LHC experiments were used in the global
fit. The same values of the QCD coupling and heavy-quark masses as
in CT10 NLO were assumed. Some results concerning CT10 NNLO PDFs were
presented at DIS2012 \cite{PNslides} and will be described here.
A longer paper on CT10 NNLO is in preparation. The CT10 NNLO PDFs
are now available in the LHAPDF library.

\begin{wrapfigure}{r}{0.4\columnwidth}%
\includegraphics[width=0.4\columnwidth]{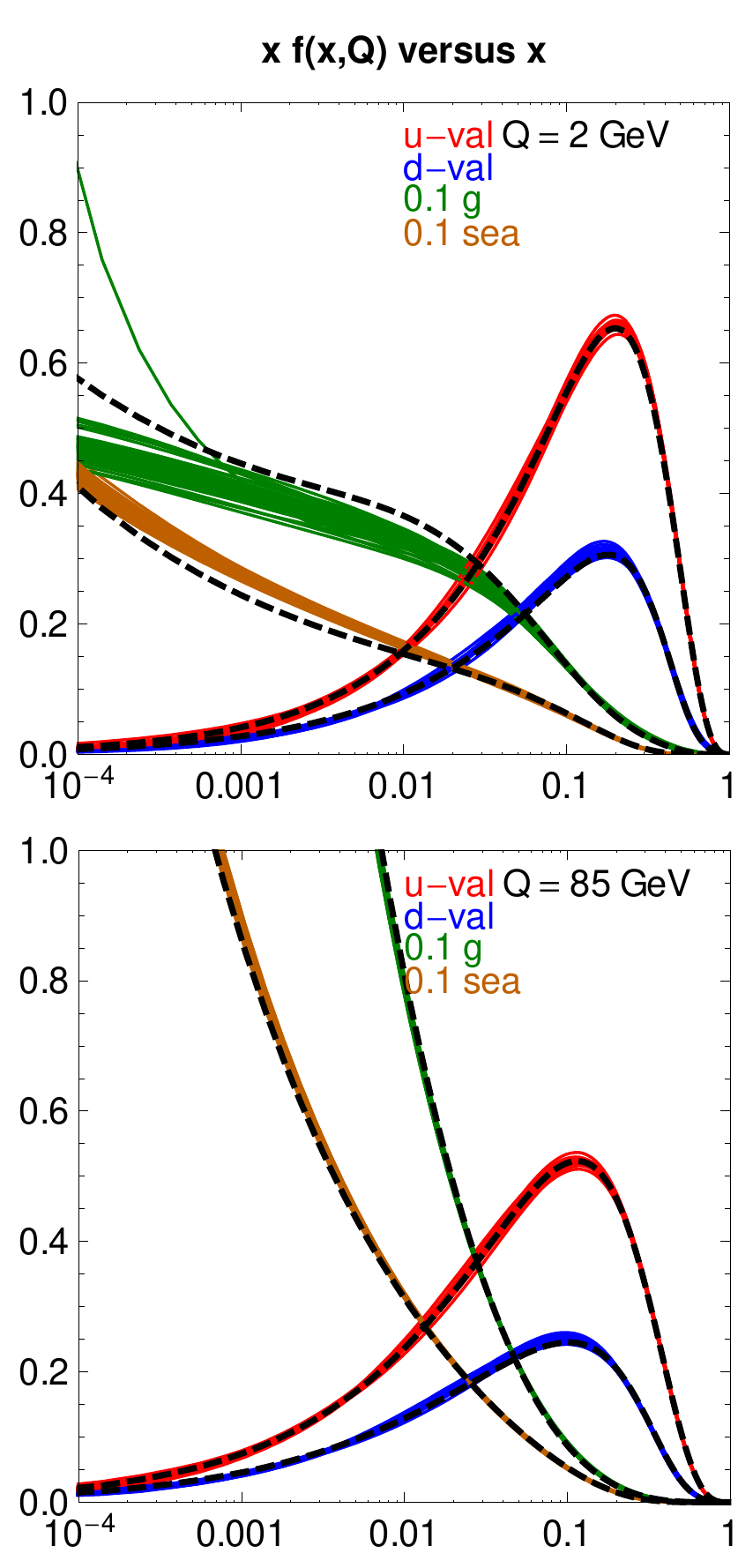} \caption{CT10 NNLO (solid color) and NLO (dashed) parton distribution functions.
\label{Fig:fig1}}
\end{wrapfigure}%

In the near future, a new release of NLO and NNLO PDFs, named CT12,
will include available data from LHC experiments. Some preliminary
results of the CT12 analysis were also presented at DIS2012 \cite{PNslides}.

\textbf{Selection of data.} At NLO, the main distinction between CT10
and CT10W sets concerns the inclusion of the D0 Run-2 $W$ electron
and muon asymmetry \cite{D0Wasy}, $A_{\ell}(y_{\ell})$, that constrains
the behavior of the ratio $d(x,Q/u(x,Q)$ at $x>0.1$. The CT10 NLO
set does not include the D0 Run-2 $A_{\ell}$\textsl{ }\textsl{\emph{data,
while the CT10W NLO set includes 4 $p_{T\ell}$ bins of $A_{\ell}$.}}
The CT10 NNLO analysis includes all data sets that were used in the
NLO fits, with the exception of the Tevatron Run-1 inclusive jet cross
sections \cite{TevJet1} that have been superceded by more precise
Tevatron Run-2 jet cross sections \cite{TevJet2}; and the D0 Run-2
$A_{\ell}$ data sets, of which only most inclusive (best understood)
bins of $p_{T\ell}$ are included in both the electron and muon channel.\textsl{\emph{
Since CT10 NNLO includes only a part of the D0 $A_{\ell}$ data that
distinguishes between CT10 NLO and CT10W NLO, it can be treated as
a counterpart of either the CT10 NLO or CT10W NLO PDF set. }}

\textbf{Overview of the PDFs.} Figure \ref{Fig:fig1} gives an overview
of the CT10 NNLO PDFs. Four PDFs are shown: $u_{{\rm valence}}(x,Q)=(u-\overline{u})(x,Q)$;
$d_{{\rm valence}}(x,Q)=(d-\overline{d})(x,Q)$; $g(x,Q)$; and $q_{sea}(x,Q)=2(\overline{d}+\overline{u}+\overline{s})(x,Q)$.
The vertical axis is $x\, f(x,Q)$. The CT10 NNLO PDFs are illustrated
by plotting all the error PDFs; hence the figure shows not only the
central fit but also the uncertainty ranges. The dashed curves are
the central-fit CT10 NLO PDFs. 

Both NLO and NNLO fits have about the same $\chi^{2}/N_{pt}\approx1.1$
for $N_{pt}=2700$ data points. Slide 5 in Ref.~\cite{PNslides}
shows a more complete comparison of CT10 NNLO to CT10W NLO, for $Q=2$\,GeV
and for thre parton flavors, $g,$ $u$, and $\bar{u}$. The various
PDFs are plotted as a ratio to the central CT10W NLO. The curves are
the ratios of the central CT10 NNLO to CT10W NLO. The shaded regions
are the \emph{error bands} for the PDFs (both NLO and NNLO). The central
NNLO PDFs differ from the central NLO PDFs, but the difference is
comparable in size to the error bands. The error band for NNLO is
slightly smaller than for NLO.

Compared to CT10W NLO, the NNLO PDF set at a small scale $Q$ has
a suppressed gluon and increased sea quarks at $x<10^{-2}$, reduced
$g(x,Q)$ and $d(x,Q)$ at $x>0.1,$ and very different charm and
bottom PDFs (slide 6 in \cite{PNslides}). The reduction in $g(x,Q)$
Compared to MSTW'08 NNLO, the central CT10 NNLO gluon PDF is somewhat
harder at $x<10^{-3}$ and $x=0.1-0.5,$ and softer at $x>0.5$ (slide
9 in \cite{PNslides}). The strangeness PDF is larger at $x\sim10^{-2}$
in CT10 NNLO than in MSTW'08 NLO, producing a good agreement with
the ATLAS measurement of the $\bar{s}(x)/\bar{u}(x)$ at this $x$
value. 

\begin{table}[!htb]
\begin{centering}
\begin{tabular}{|c|c|c|c|}
\hline 
Boson/collider & CT10 NLO & CT10 NNLO & MSTW'08 NNLO \tabularnewline
\hline 
$W^{+}$ LHC14 (nb)  & $12.2\pm0.5$  & $12.7\pm0.5$  & $12.4\pm0.2$ \tabularnewline
\hline 
$W^{+}$ LHC7 (nb)  & $6.0\pm0.2$  & $6.3\pm0.2$  & $6.2\pm0.1$ \tabularnewline
\hline 
$W^{+}$ Tevatron (nb)  & $1.35\pm0.05$  & $1.38\pm0.05$  & $1.38\pm0.02$ \tabularnewline
\hline 
$W^{-}$ LHC'14 (nb)  & $8.9\pm0.4$  & $9.4\pm0.4$  & $9.3\pm0.2$ \tabularnewline
\hline 
$W^{-}$ LHC'7 (nb)  & $4.10\pm0.15$  & $4.29\pm0.16$  & $4.31\pm0.07$ \tabularnewline
\hline 
$Z$ LHC14 (nb)  & $2.07\pm0.08$  & $2.17\pm0.08$  & $2.13\pm0.03$ \tabularnewline
\hline 
$Z$ LHC7 (nb)  & $0.96\pm0.03$  & $1.00\pm0.03$  & $0.99\pm0.02$ \tabularnewline
\hline 
$Z$ Tevatron (pb)  & $260\pm9$  & $263\pm8$  & $261\pm5$ \tabularnewline
\hline 
$H_{SM}^{0}$ LHC14 (pb)  & $101\pm9$  & $99\pm8$  & $102\pm7$ \tabularnewline
\hline 
$H_{SM}^{0}$ LHC7 (pb)  & $31.2\pm1.9$  & $29.7\pm1.7$  & $29.8\pm1.3$ \tabularnewline
\hline 
$H_{SM}^{0}$ Tevatron (pb)  & $1.77\pm0.12$  & $1.77\pm0.12$  & $1.80\pm0.11$ \tabularnewline
\hline
\end{tabular}
\par\end{centering}

\caption{Total cross sections for production of electroweak bosons.}

\label{TAB:tab1} %
\end{table}

\vspace{-20pt}%
\begin{figure}[h]
\includegraphics[width=0.3\textwidth]{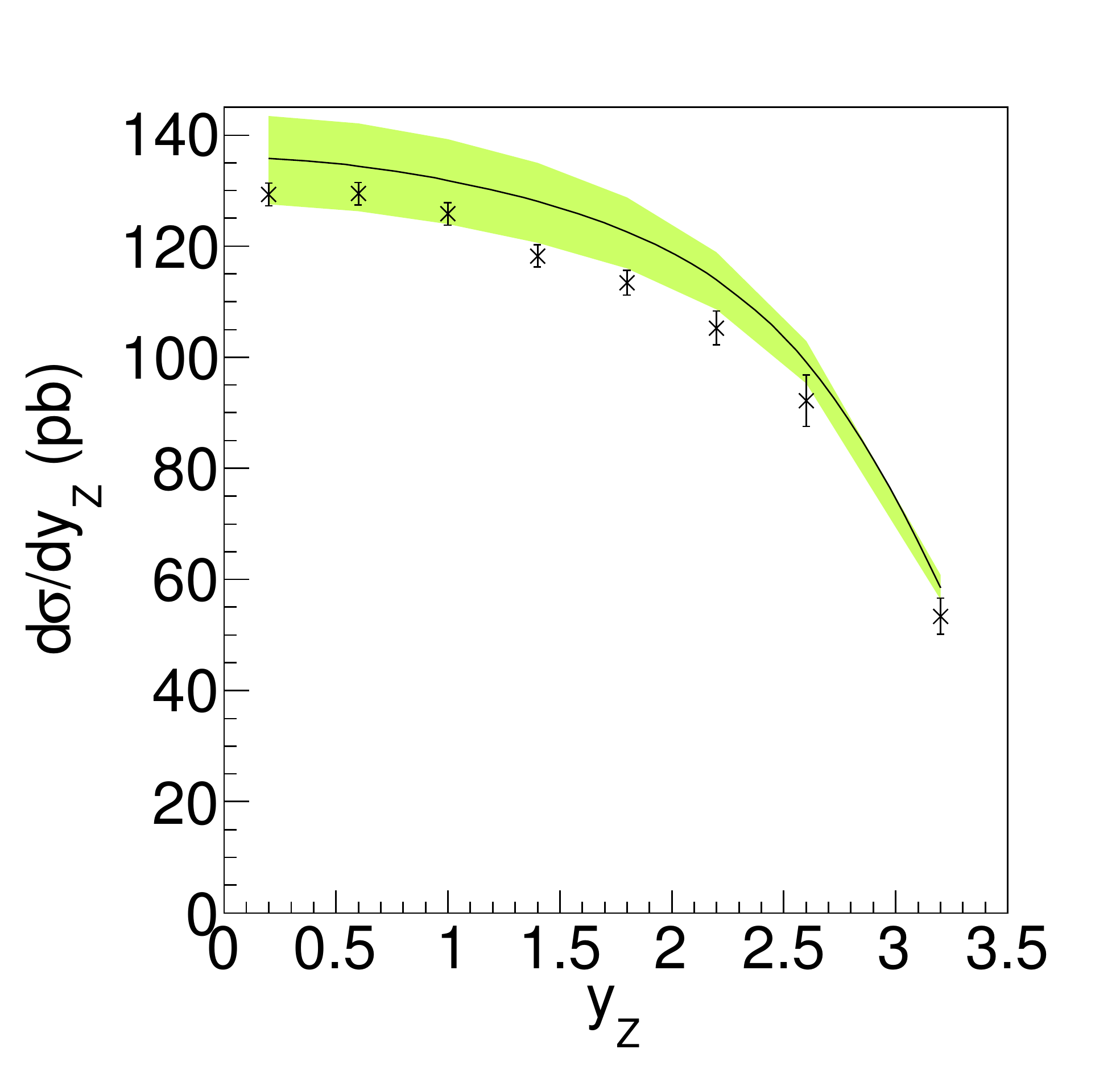}\includegraphics[width=0.3\textwidth]{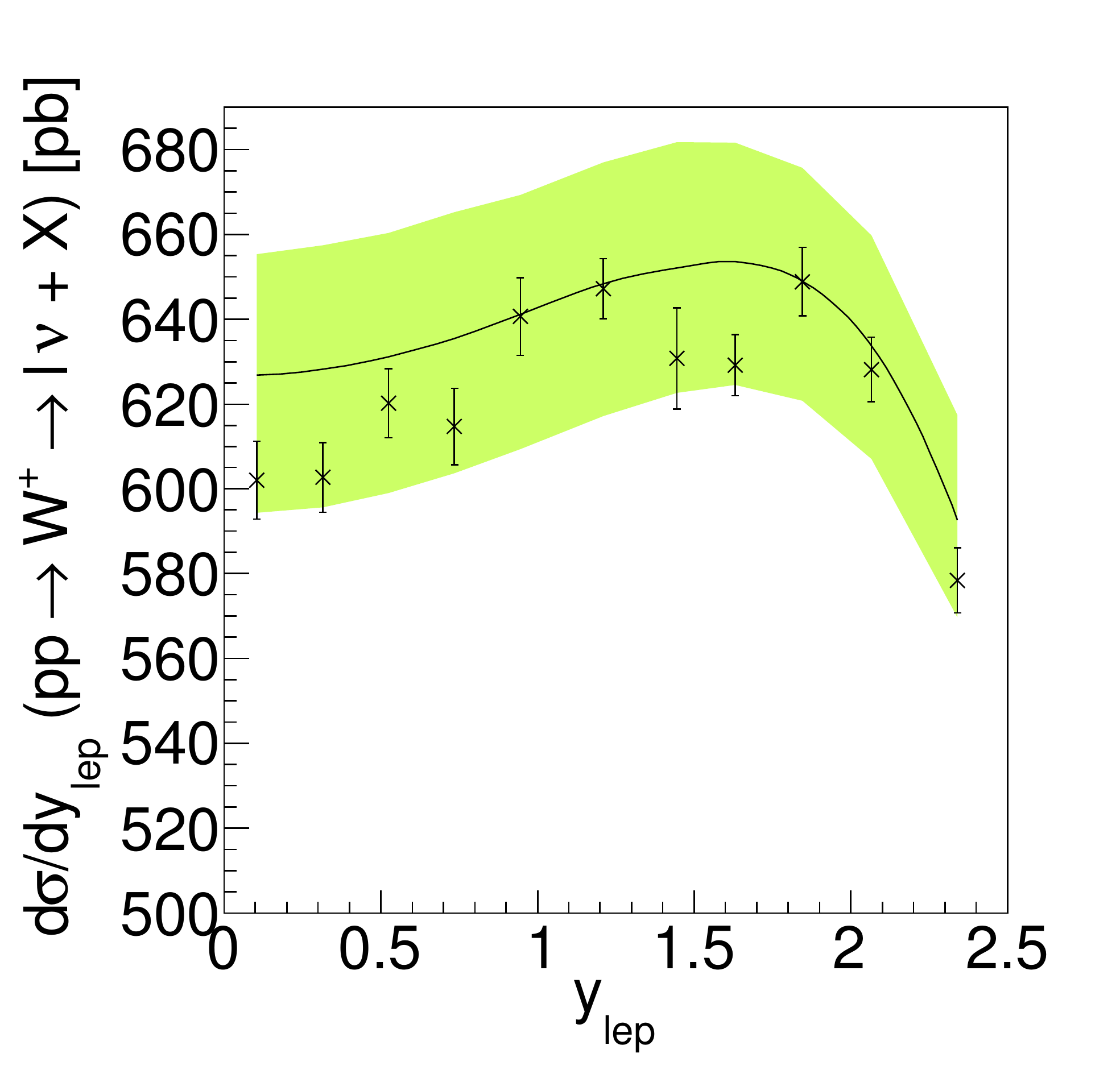}\includegraphics[width=0.3\textwidth]{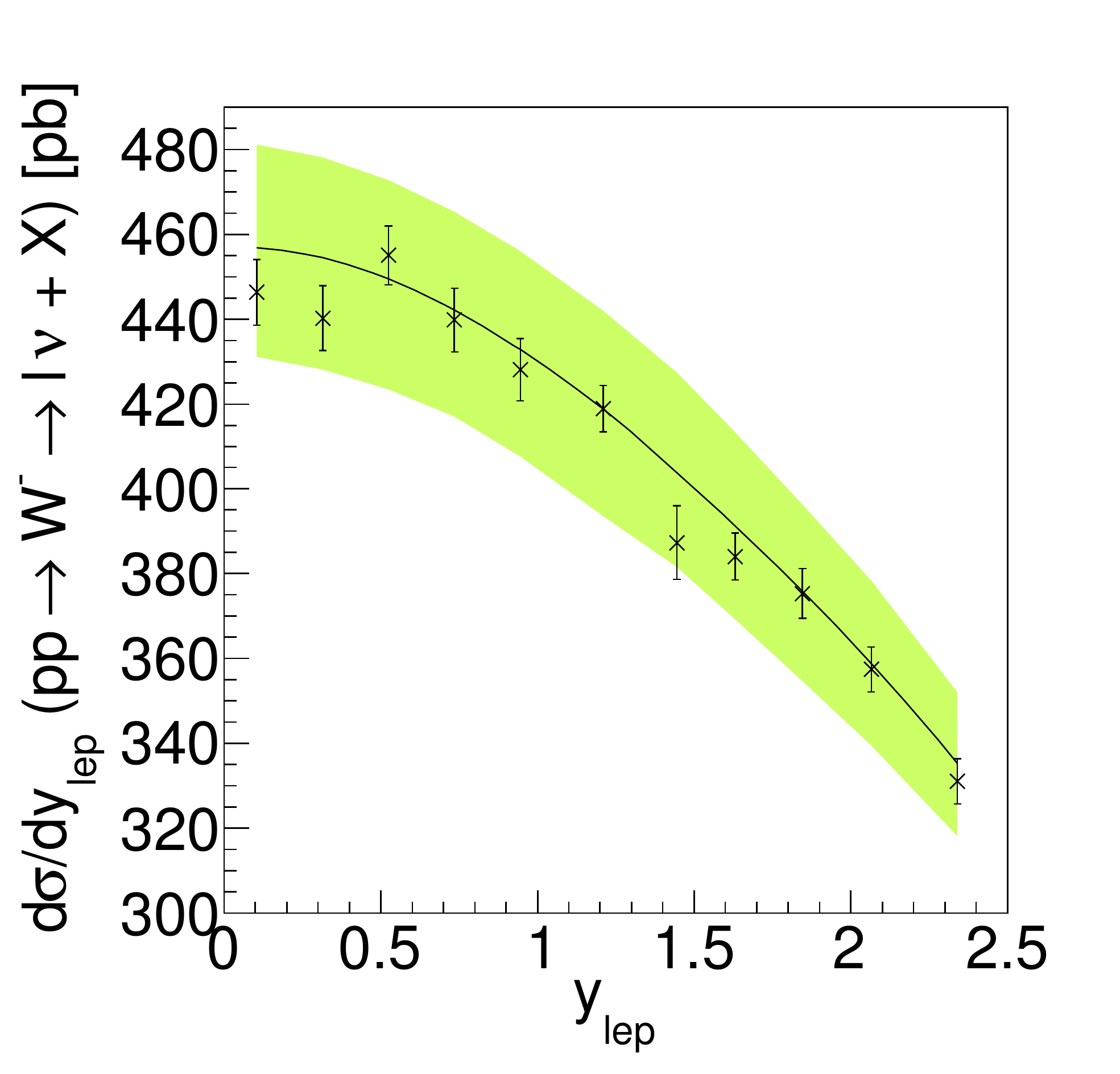}

\caption{Comparisons of ATLAS data with ResBos predictions for $Z^{0}$ and
$W^{\pm}$ lepton rapidity distributions. \label{Fig:fig2}}
\end{figure}

\textbf{Predictions for the LHC.} In a future paper we will provide
detailed comparisons of theory and data, where the theory is calculated
from the CT10 NNLO PDFs. Here we collect some representative cross
sections for the hadron colliders. Table \ref{TAB:tab1} compares
predictions for total cross sections for $W$, $Z$ and Higgs boson
production via gluon fusion (with Higgs mass of 125 GeV) at the Tevatron
and the LHC (with $\sqrt{s}=$7 and 14 TeV). The comparison is between
CT10 NLO, CT10 NNLO, and MSTW'08 NNLO. The CT10 NNLO central PDF increases
the total cross sections by a few percent compared with CT10 NLO accuracy
and is close to MSTW'08. Theoretical uncertainties from alternative
PDF sets for CT10 NNLO are similar to those for CT10, and in W/Z production
they are about twice as those for MSTW'08.

Fig.~\ref{Fig:fig2} shows the comparison of ATLAS data \cite{Aad:2011dm}
with ResBos \cite{ResBos} predictions for $Z$ and $W$-lepton rapidity
distributions at the LHC ($\sqrt{s}=7$ TeV) using CT10 NNLO PDFs.
Theoretical uncertainty bands were calculated using the error PDF
sets. The ResBos prediction of $Z$ and $W^{+}$-lepton rapidity distribution,
using the central PDF set, is higher than ATLAS data by a few percent.
However, for $W^{-}$-lepton rapidity distribution, the ResBos prediction
is more consistent with ATLAS data. It is expected that these data
could further refine the PDFs at the NNLO accuracy.

\addtolength{\textheight}{15pt}%
\begin{figure}[!tb]
\includegraphics[width=0.43\textwidth]{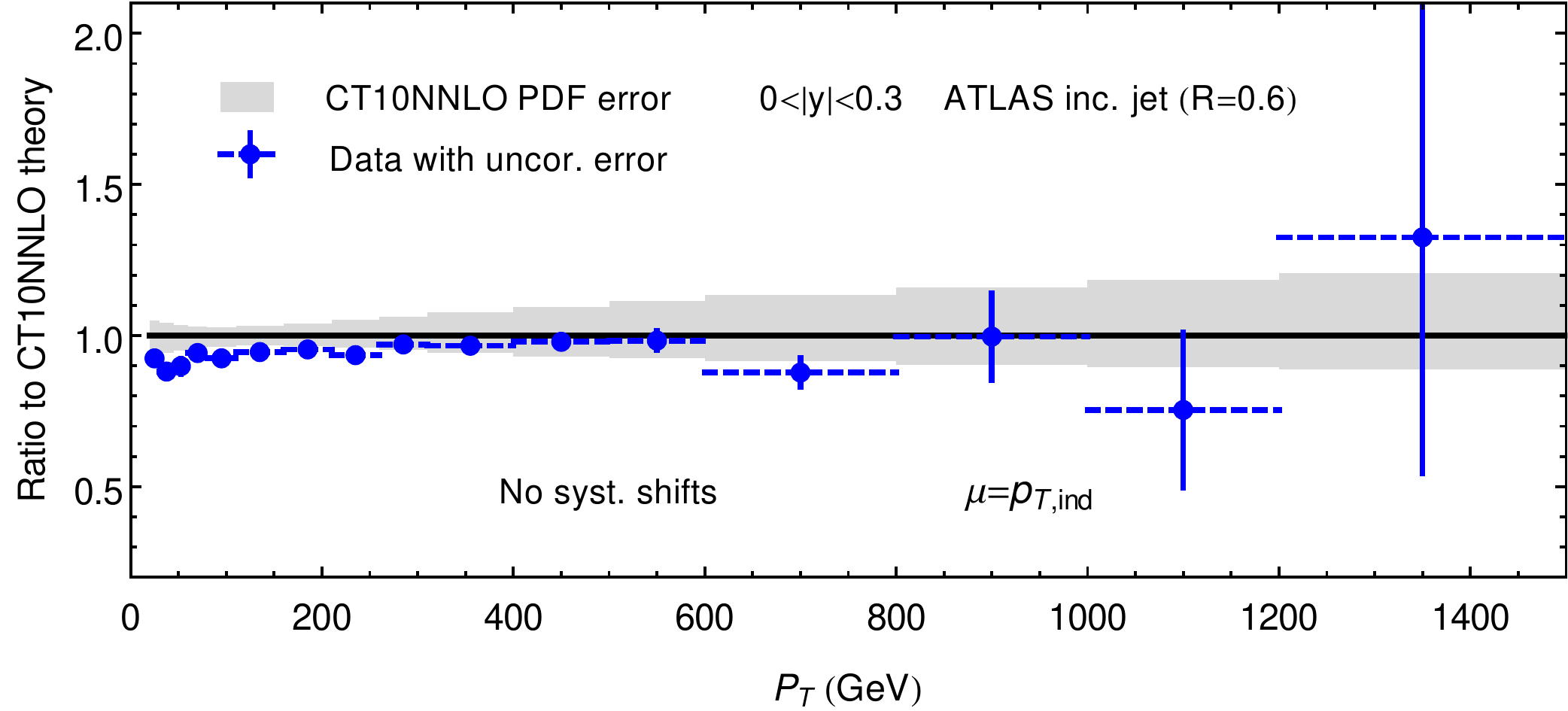} ~~\includegraphics[width=0.43\textwidth]{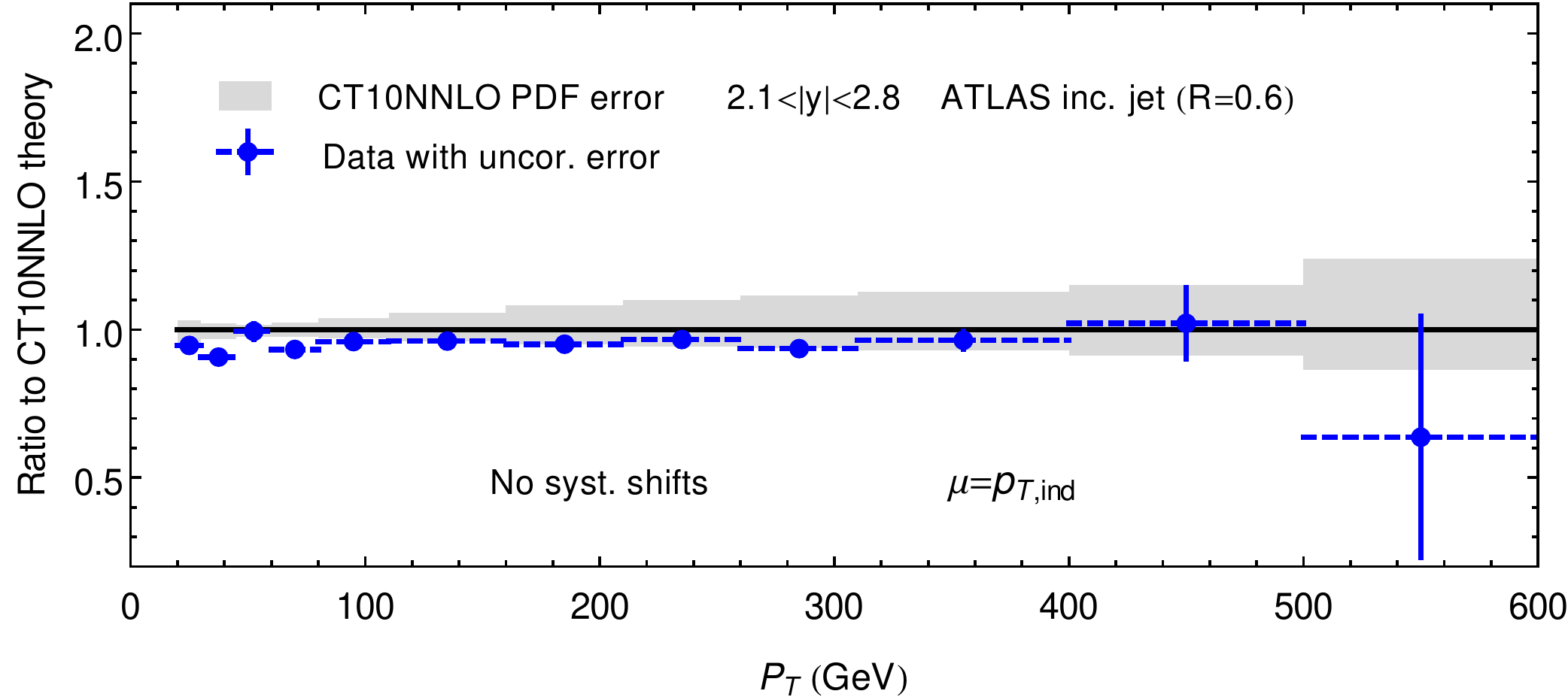} 

\includegraphics[width=0.43\textwidth]{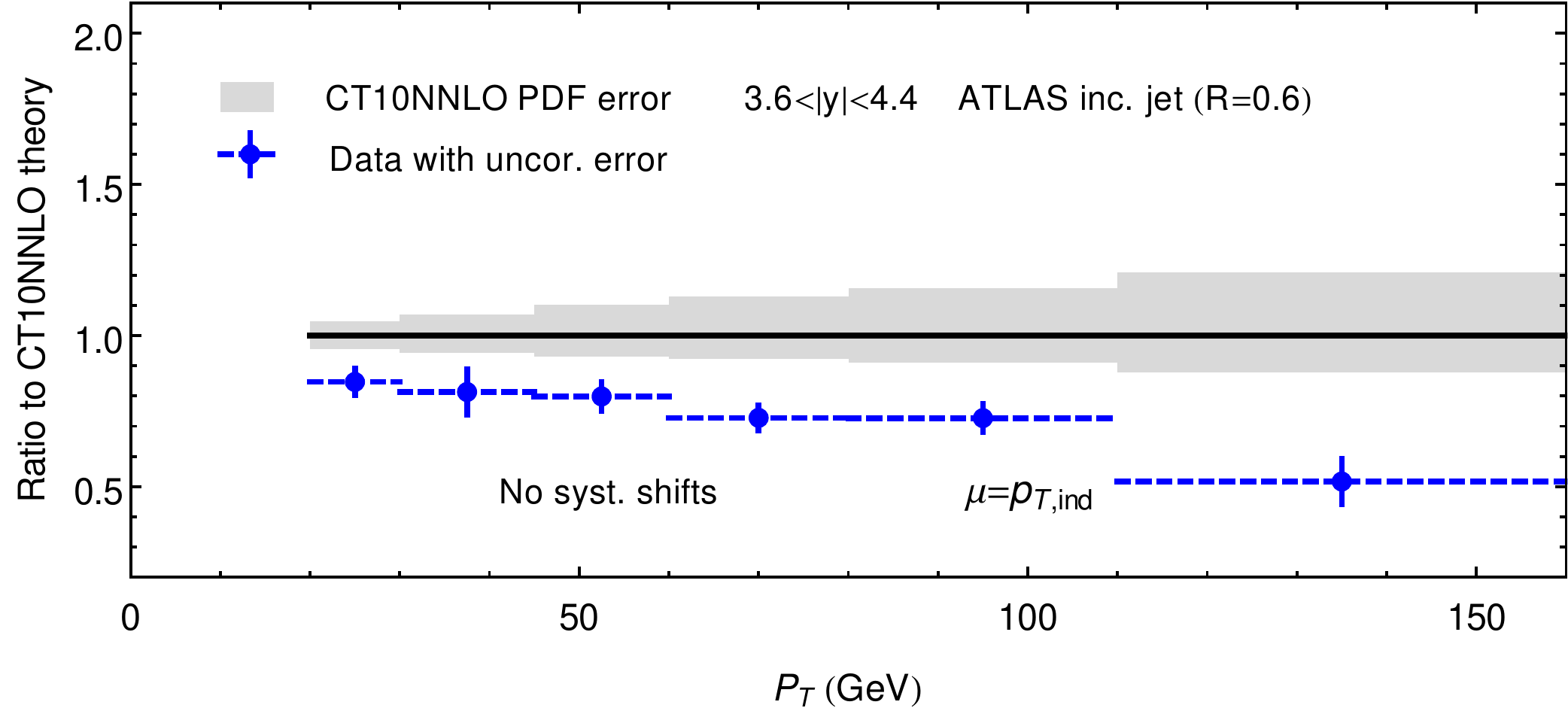}~~~%
\begin{minipage}[t]{0.5\columnwidth}%
\vspace{-80pt}\caption{Comparison of ATLAS data for inclusive jet $p_{T}$ distribution with
a theoretical prediction using CT10 NNLO. \label{Fig:fig3}}
\end{minipage}%
\end{figure}

Fig.~\ref{Fig:fig3} compares the ATLAS data for inclusive jet transverse
momentum distribution with theoretical predictions based on the NLO
matrix elements and CT10 NNLO PDFs. They agree well even without including
the systematic shifts, except for the large rapidity region. After
accounting for the systematic shifts, the reduced $\chi^{2}$ is 0.78
for the measurement with R=0.4 and 0.76 for the one with R=0.6. The
effect of the LHC data on the PDFs will be explored in the CT12 analysis.

This work was supported by the U.S. DOE Early Career Research Award
DE-SC0003870 and by the U.S. NSF under grant No.~PHY-0855561.

\raggedright \begin{footnotesize} 

\end{footnotesize}
\end{document}